\input{aipcheck}

\documentclass[
    ,final            
  ]
  {aipproc}

\layoutstyle{6x9}

\def\eslt{E_T^{\rm miss}}

\def\to{\rightarrow}

\def\bi{\begin{itemize}}
 \def\ei{\end{itemize}}

\def\c1p{C1^\prime}
\def\ta{\tilde a}
\def\tG{\tilde G}

\def\ta{\tilde a}

\def\tg{\tilde g}
\def\tnu{\tilde\nu}
\def\tell{\tilde\ell}
\def\tq{\tilde q}

\def\tw{\tilde\chi}

\def\tz{\tilde\chi^0}
\def\alt{\stackrel{<}{\sim}}
\def\agt{\stackrel{>}{\sim}}
\def\be{\begin{equation}}  
\def\ee{\end{equation}}  
\def\bea{\begin{eqnarray}}  
\def\eea{\end{eqnarray}}

\newcommand\prd[3]{{\it Phys.\ Rev.\ }{\bf D #1} (#2) #3}

\newcommand\prl[3]{{\it Phys.\ Rev.\ Lett.\ }{\bf #1} (#2) #3}
\newcommand\plb[3]{{\it Phys.\ Lett.\ }{\bf B #1} (#2) #3}
\newcommand\jhep[3]{{\it J. High Energy Phys.\ }{\bf #1} (#2) #3}

\newcommand\npb[3]{{\it Nucl.\ Phys.\ }{\bf B #1} (#2) #3}

\newcommand\ptp[3]{{\it Prog.\ Theor.\ Phys.\ }{\bf #1} (#2) #3}
\newcommand\zpc[3]{{\it Z.\ Physik }{\bf C #1} (#2) #3}

\newcommand\jphg[3]{{\it J. Phys.\ }{\bf G #1} (#2) #3}
\begin{document}

\title{Leptonic signatures for SUSY at the LHC}

\classification{11.30.Pb,12.60.Jv,14.80.Ly}
\keywords      {Supersymmetry phenomenology, collider physics}

\author{Howard Baer
\footnote{Plenary talk presented at the 
17th International Conference on Supersymmetry and the 
Unification of Fundamental Interactions (SUSY09) at 
Northeastern University, Boston, MA, 5-10 June, 2009.}
}{
  address={Homer L. Dodge Department of Physics,
University of Oklahoma, Norman, OK 73019, USA}
}

\begin{abstract}
Most models of weak scale supersymmetry (SUSY) predict observable rates for 
production of SUSY matter at the CERN LHC. The SUSY collider events are
expected to be rich in jets, isolated (and non-isolated) leptons and 
missing $E_T$. After first discussing the merits of mixed axion/axino vs.
neutralino cold dark matter in SUSY models, I then 
survey prospects for detecting SUSY matter at the LHC via 
leptonic signatures. In the paradigm mSUGRA model, 
cascade decays of gluinos and squarks should yield
high rates for multi-jet plus multi-lepton events, allowing 
values of $m_{\tg}\sim 3$ (1.8) TeV to be probed for $m_{\tq}\simeq m_{\tg}$
($m_{\tq}\gg m_{\tg}$) with 100 fb$^{-1}$ of integrated luminosity.
Direct production of gauginos and sleptons should also be possible in limited
regions of parameter space.
Even in the first year of LHC running, observable signals in multi-muon 
plus jets channel (without cutting on missing $E_T$) can occur for interesting
ranges of parameters. The highly motivated Yukawa unified SUSY models-- 
where the dark matter is expected to be of mixed axion/axino type--
should likely be testable in the first year of LHC running due to 
large rates for gluino pair production followed by cascade decays.
\end{abstract}

\maketitle


\section{Supersymmetric models}

Particle physics models including weak scale supersymmetry (SUSY) are highly motivated 
both from the theoretical as well as the experimental point of view\cite{wss}.
On the theory side, SUSY stabilizes the Higgs sector, and allows one to 
extrapolate physics safely to very high energy scales. On the experimental side,
the most impressive argument comes from extrapolating the measured values of the 
three Standard Model (SM) gauge couplings from the weak scale to the GUT scale. The celebrated
unification of gauge couplings at $M_{GUT}\simeq 2\times 10^{16}$ GeV seems to
indicate that 1. the Minimal Supersymmetric Standard Model (MSSM), or MSSM plus gauge singlets
(or extra $SU(5)$ multiplets), is the correct effective field theory all the way up to
$M_{GUT}$ and that 2. the unification certainly looks GUT-like, and that a SUSY GUT theory may be
the correct effective field theory around $Q\agt M_{GUT}$. Of the various GUT theories,
$SO(10)$ stands out in that it unifies not only the three SM forces, but also all the particles
of each SM generation (into the 16 dimensional spinor of $SO(10)$). In the simplest $SO(10)$
SUSY GUT models, Yukawa couplings of the third generation are also expected to unify.

There are a host of SUSY models which are consistent with gauge coupling unification.
Some of them I list below:
\begin{itemize}
\item gauge-mediated SUSY breaking (GMSB),
\item anomaly-mediated SUSY breaking (AMSB),
\bi
\item mixed moduli-AMSB (mirage unification models)
\item hypercharged AMSB
\item deflected AMSB
\ei
\item gravity -mediated SUSY breaking models\cite{sugra},
\bi
\item mSUGRA (also known as CMSSM)
\item one and two parameter non-universal Higgs models (NUHM1,NUHM2),
\item non-universal gaugino masses in various guises
\item normal scalar mass hierarchy (broken generations) with $m_0(1,2)> m_0(3)$,
\item compressed SUSY
\item split SUSY, pMSSM, NMSSM, 
\item $\cdots$ .
\ei
\end{itemize}

Lately, gravity-mediated SUSY breaking models seem most popular because they
can easily accommodate SUSY breaking via supergravity effects, and seem to
most easily accommodate cold dark matter (CDM) in the universe. 
Of the gravity-mediated models, most work has been done on the mSUGRA (or CMSSM) model. 
Whether it is right or wrong, it is at least simple, consistent with all data, 
and exhibits many intriguing features which might be observable in the next round of collider 
and dark matter experiments. So most of the results I show will come from that model.
At the end, I will coment on Yukawa-unified models, which only seem to occur when non-universality 
of soft SUSY breaking terms is allowed, as in the NUHM2 model.

\subsection{Neutralino vs. axion/axino cold dark matter}

\begin{figure}
  \includegraphics[height=.3\textheight]{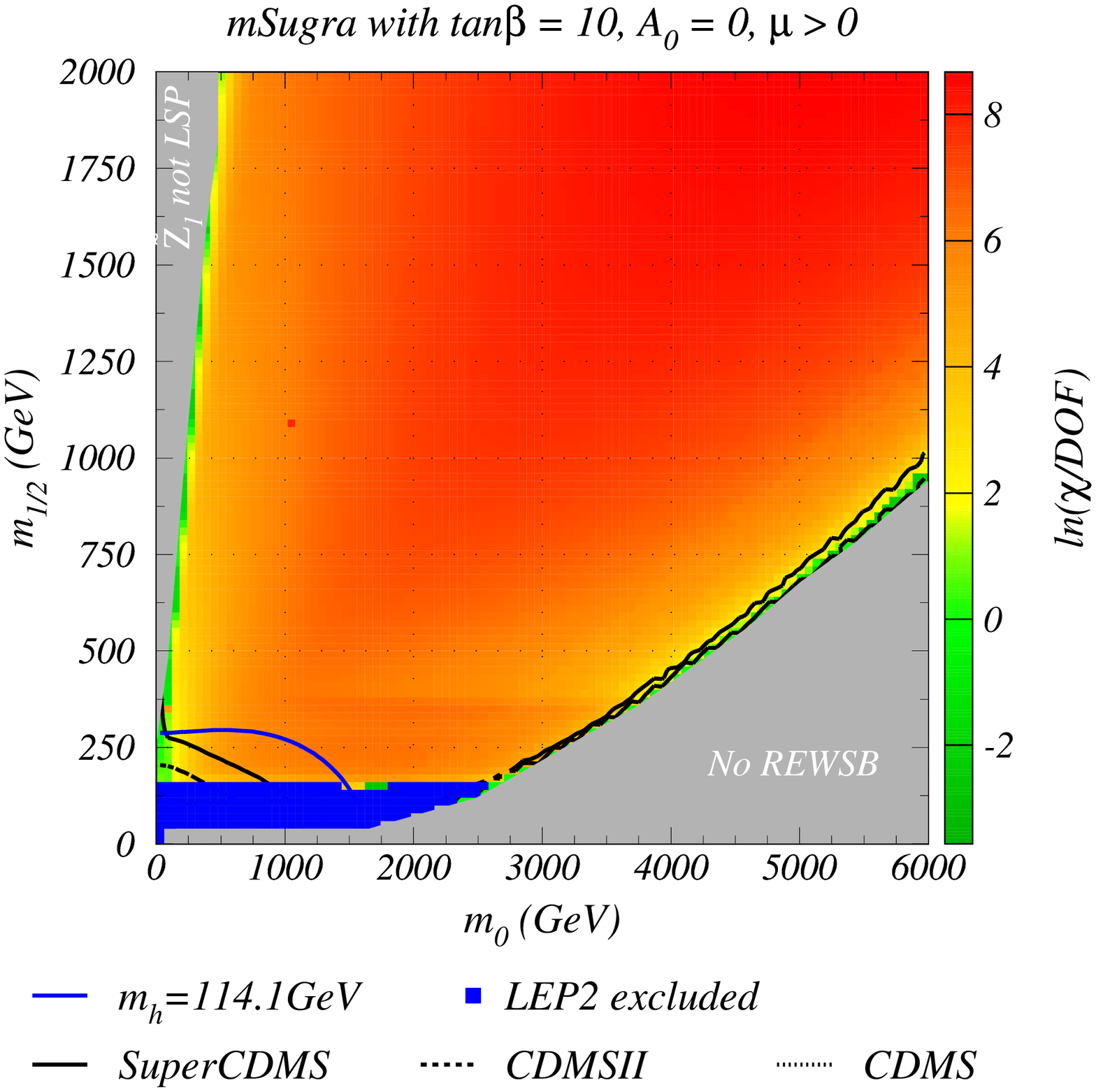}
  \includegraphics[height=.3\textheight]{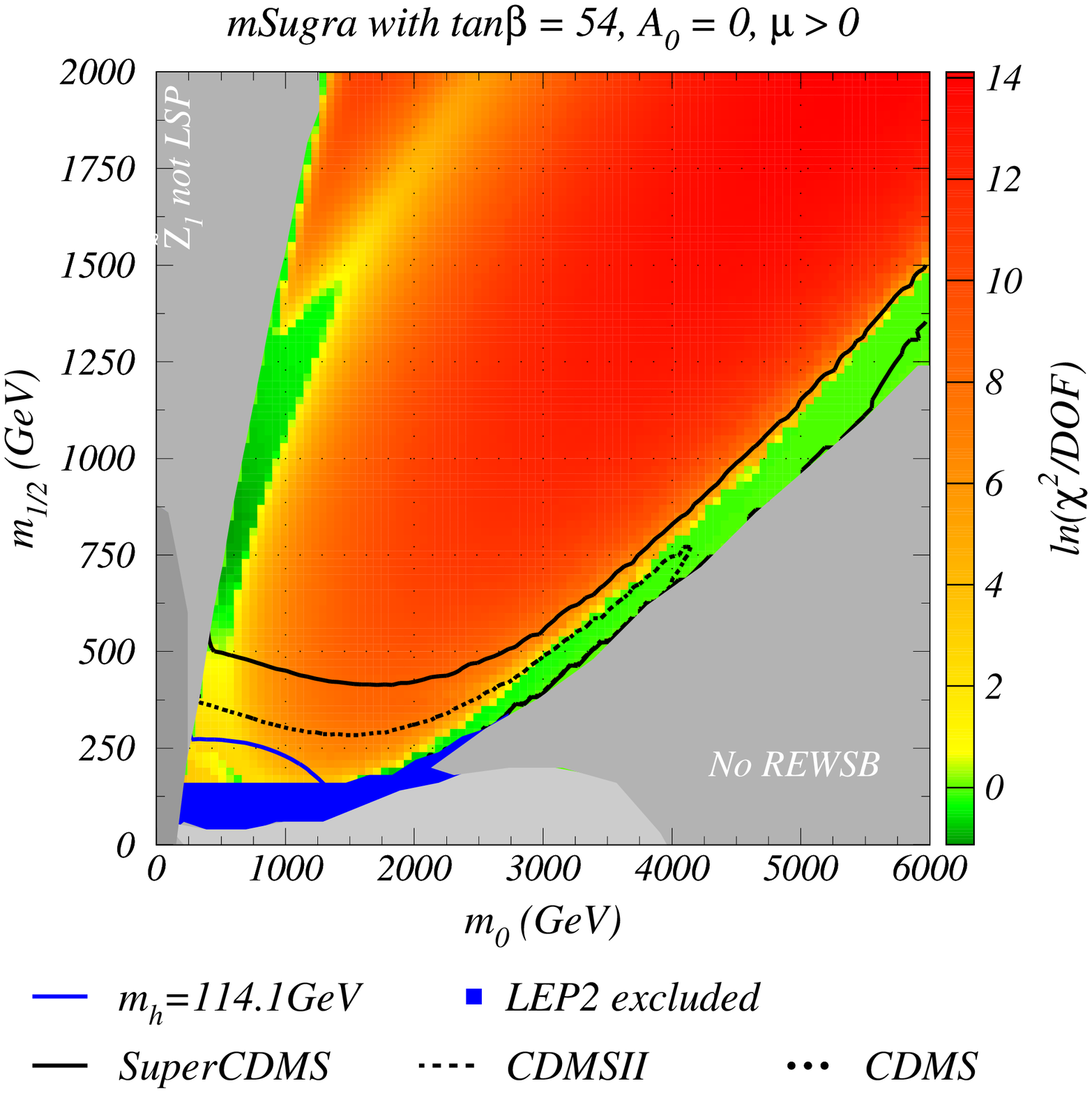}
  \caption{$\chi^2$ fit of neutralino abundance, $(g-2)_\mu$ and $BF(b\to s\gamma )$ 
to SUSY parameters (in this case, using $\tau$ data for hadronic vacuum polarization 
in $(g-2)_\mu$).}
\label{fig:bb}
\end{figure}

The lightest neutralino of SUSY, the $\tz_1$ state, is a prototypical WIMP dark matter
candidate\cite{haim}. 
The neutralino relic abundance can be calculated in SUSY models, and is embedded
in public codes such as DarkSUSY, MicroMegas and IsaReD (the latter a part of the Isajet
event generator). Several groups have been fitting the dark matter density, $BF(b\to s\gamma )$,
$(g-2)_\mu$, LEP2 constraints, plus possibly other EW observables, to SUSY models. I show
here in Fig. \ref{fig:bb} results from Balazs and I from 2003\cite{bb}, 
since nothing of key importance has alterred the situation since then.

The green regions show the good fit, and it is mainly governed by the fit to the WMAP-measured
CDM density in the universe. Exhibited are 1. the stau co-annihilation region (left edge), 
2. the hyperbolic branch/focus point (HB/FP) region (right edge), 3. the $A$-resonance annihilation region
(at large $\tan\beta$ only), and 4. a bit of the light Higgs $h$-resonance annihilation (at low $m_{1/2}$) 
and 5. the so-called
bulk annihilation region (at low $m_0$ and low $m_{1/2}$, now largely excluded by LEP2).
Most of the parameter space gives too much dark matter, and so is excluded. So much for the 
WIMP miracle! Neutralinos can make up the bulk of dark matter only under very fine-tuned conditions.

SUGRA based models suffer another important constraint: the gravitino problem.
Gravitinos can be produced at large rates in the early universe. If gravitinos are
heavier than the other SUSY particles, then they decay into them with late-time decays, which can
disrupt the successful predictions of Big Bang Nucleosynthesis (BBN). To avoid this\cite{kkmy},
one must have, roughly, that $m_{\tG}\agt 5$ TeV or the re-heat temperature of the universe
$T_R\alt 10^5$ GeV (which conflicts with compelling baryogenesis scenarios like leptogenesis).
One might try to avoid the gravitino problem by making the gravitino the lightest SUSY particle (LSP).
But then thermal production of SUSY particles, followed by late-time decays to gravitinos, again disrupts BBN, 
unless (roughly) $m_{\tG}\alt .1-1$ GeV\cite{kkmy}.

In addition, another problem (that one neglects at one's peril) is the strong CP problem.
The compelling solution here is the original Peccei-Quinn-Weinberg-Wilczek solution\cite{pqww}, 
which implies existence of an axion particle $a$. Cosmology constrains the PQ breaking scale
$10^{8}\alt f_a/N\alt 10^{12}$ GeV, which means $10^{-6}\alt m_a\alt 10^{-3}$ eV.
Since we are in supersymmetry, the axion must be accompanied by a spin-$1\over 2$ super-partner the axino $\ta$\cite{rtw}.
The axino mass is relatively unconstrained: it can lie anywhere between the keV and multi-GeV range\cite{ckkr}.
If $m_{\ta}<m_{\tz_1}$, then the $\ta$ can be the LSP. In this case, dark matter can consist
of a mixture of cold axions produced via vacuum mis-alignment\cite{as}, thermally produced axinos (whose abundance
depends on $T_R$)\cite{bs}, and non-thermally produced axinos arising from neutralino decay\cite{ckkr}. 

One can invoke the PQWW strong CP solution within the context of mSUGRA. 
In the case that the $\tz_1$ is the lightest MSSM particle, then it will decay $\tz_1\to \ta\gamma$
(or possibly other modes) with a lifetime of order 1 second or less. Thus, it avoids the BBN problem
(as long as $\tG$ is heavy), but neutralinos will still give rise to $\eslt$ at colliders.
 
\begin{figure}
  \includegraphics[height=.3\textheight]{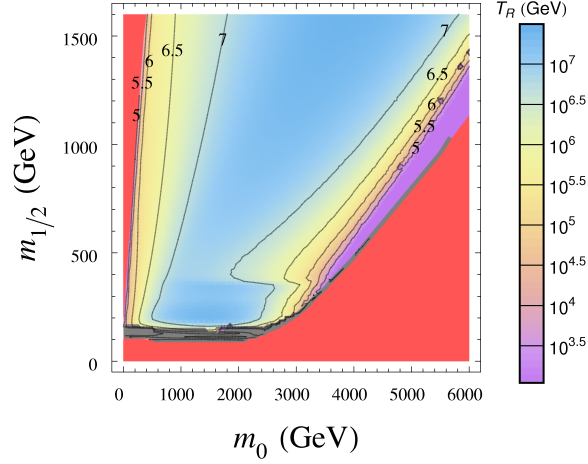}
  \caption{Contours of $T_R$ needed to generate the WMAP measured CDM abundance using
mixed axion/axino (but mainly axion) CDM in the mSUGRA model.
}
\label{fig:axdm}
\end{figure}

For a given hypothesis
of $f_a/N$ and $m_{\ta}$, the WMAP-measured CDM abundance allows a calculation of $T_R$. Contours
of $T_R$ are shown in Fig. \ref{fig:axdm}\cite{bbs}. Everywhere in the plane one gets the correct WMAP
abundance. The blue regions have $T_R>10^7$ GeV, which at least allows for non-thermal
leptogenesis\cite{ntlepto}. 
The relevance of this plot for leptonic signatures at the LHC is that the entire LEP2-allowed
parameter space of mSUGRA is also CDM-allowed, in the case of mixed axion/axino CDM: one should not
focus just on the special neutralino DM-allowed regions for LHC SUSY signatures.
Now on to leptonic LHC signatures!

\section{Slepton pair production}

Direct production of sleptons can take place at LHC via the Drell-Yan mechanism:
$s$-channel $\gamma$ and $Z$ exchange leads to $\tell_L\bar{\tell}_L$, 
$\tell_R\bar{\tell}_R$, $\tnu_\ell\bar{\tnu}_\ell$ production, while $s$-channel $W$
exchange leads to $\tell\tnu_\ell$ production. The $\tnu_\ell$ may or may not decay to
visible states. These reactions were investigated a long time ago\cite{bcpt1},
and the best signature was to look for $\ell^+\ell^- +\eslt$ final states arising from
slepton pair production. Requiring $p_T(\ell)>40$ GeV, $\eslt >100$ GeV, a central jet veto
and $\delta\phi (\ell^+\ell^- )<90^\circ$ gave observable signals against $W^+W^-$, 
$Z\to\tau^+\tau^-$ and $t\bar{t}$ backgrounds for $m_{\tell}\alt 350$ GeV for 
10 fb$^{-1}$ of integrated luminosity at LHC with $\sqrt{s}\sim 14$ TeV. 
Similar studies were performed by Denegri, Majerotto and Rurua\cite{dr}.

\section{Clean trileptons from chargino-neutralino production}

A clean (jet-free) trilepton signature can arise from $pp\to \tw_1\tz_2 X$ production,
followed by $\tw_1\to\ell\nu_\ell\tz_1$ and $\tz_2\to \ell\bar{\ell}\tz_1$ decay\cite{lhc3l}.
In fact, this SUSY production cross section can be the dominant one at LHC in the case where
squarks and sleptons have mass greater than about a TeV. In the case of the clean $3\ell +\eslt$
signal, we require each lepton to obey a ``cone'' isolation requirement to reject leptons arising from heavy flavor decay, and then require $p_T(\ell )>20,20,10$ GeV 
for the three hardest isolated leptons. We will also require a ``central jet veto'' (clean
trileptons), and $\eslt $ {\it less than} 100 GeV, Finally, require a leptonic $Z$ veto to
reject BG from WZ production. 
In this case, SM backgrounds are generally lower than SUSY signal in regions where the 
$\tz_2\to \ell^+\ell^-\tz_1$ branching fraction is substantial. It is substantial as long as other
two body ``spoiler'' modes such as $\tz_2\to \tz_1 h$ or $\tz_1 Z$ are closed, or there is 
not large interference in the 3-body neutralino decay which suppresses the leptonic BFs.
A virtue of this signal is that the dilepton pair from $\tz_2\to\tz_1\ell^+\ell^-$
is kinematically constrained to obey $m(\ell^+\ell^- )<m_{\tz_2}-m_{\tz_1}$ (for 3-body
decays) or a similar constraint if $\tz_2\to\ell^\pm\tell^\mp\to 3\ell$ occurs\cite{frank}.

\section{Gluino and squark cascade decays to multi-lepton plus jets states}

While slepton pair production and gaugino pair production can lead to clean multi-lepton
events as LHC, we also expect multiple isolated lepton plus multijet plus $\eslt$ events to
arise from gluino and squark production\cite{cascade}. The strong interaction processes 
$pp\to\tg\tg,\ \tq\tq,\ \tg\tq +X$ are expected to be the dominant SUSY production modes
at LHC as long as $m_{\tg},\ m_{\tq}\alt 1$ TeV. 

The $\tg$ and $\tq$ states will then undergo ``cascade decays'': possibly 
multi-step decay processes into SM particles plus the lighter superpartners until
the state containing the LSP is reached. There are of order 1000 sparticle
subprocess production reactions, and numerous decay modes of each sparticle, 
which are listed as output by programs such as IsaSUSY/IsaSUGRA, SUSYHIT and Spheno.
Thus, roughly $10^5$ sparticle $2\to n$ reactions can occur at LHC in the case of the MSSM.
The exact decay patterns are model dependent, and vary significantly around parameter space of any
model. Generally, we expect gluno and squark cascade decay events to contain
numerous high $E_T$ jets (including numerous $b$-jets and possibly $\tau$ jets, especially at
large $\tan\beta$\cite{ltanb}), numerous isolated $e$s and $\mu$s and $\eslt$. 

\begin{figure}
  \includegraphics[height=.3\textheight]{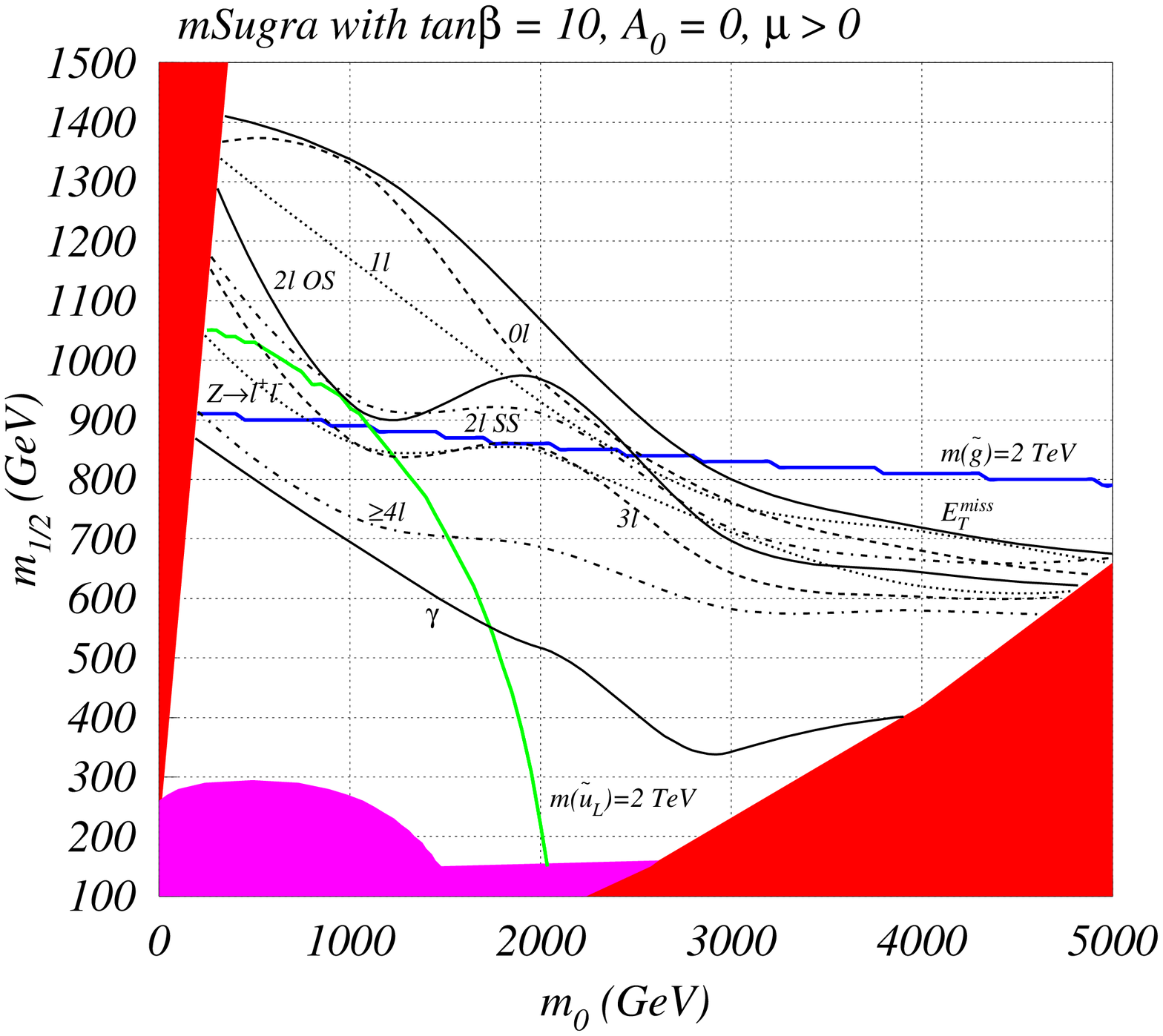}
  \includegraphics[height=.3\textheight]{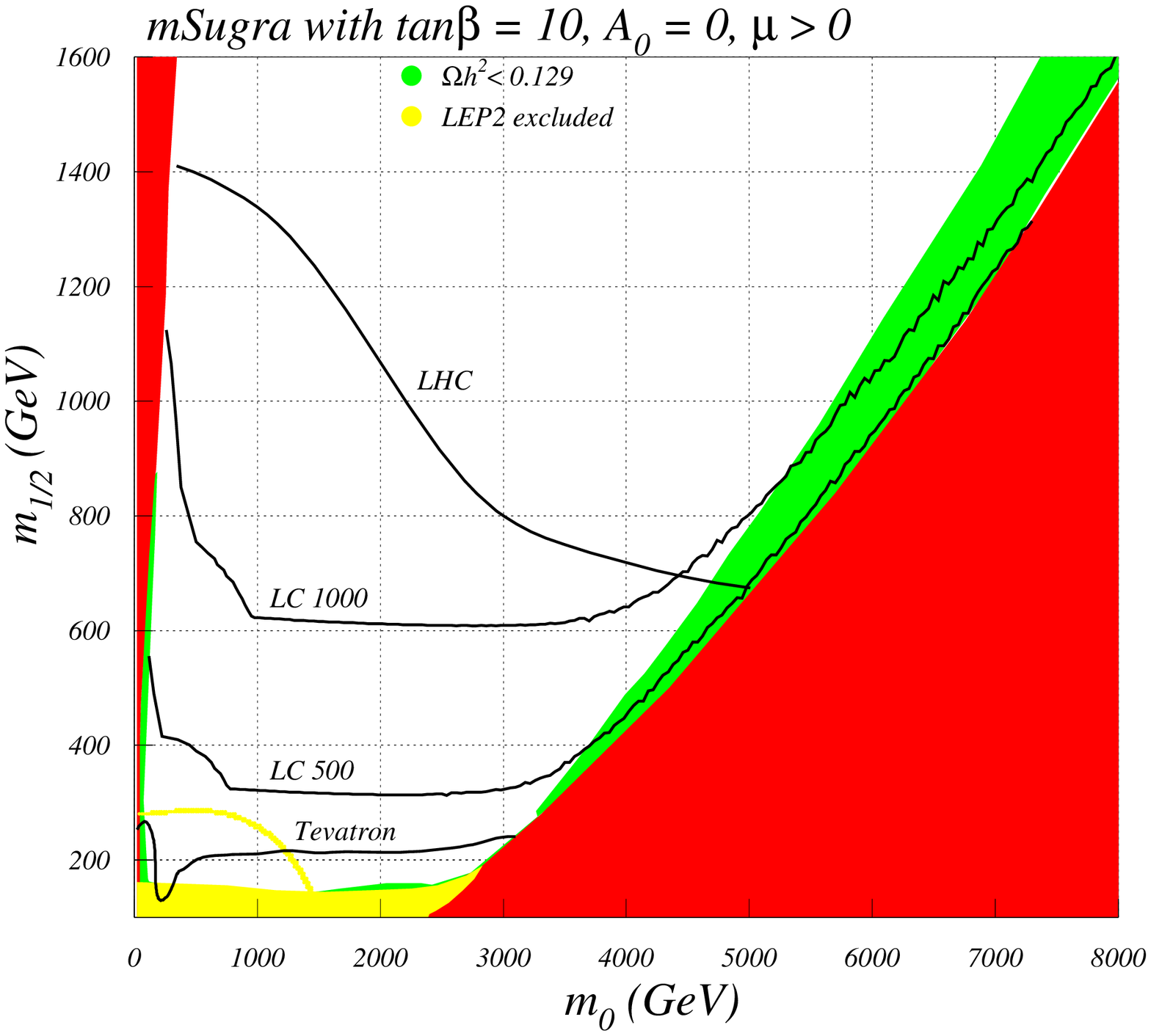}
  \caption{{\it a}). The reach of LHC for SUSY in various event topologies assuming
100 fb$^{-1}$ in the mSUGRA model. 
{\it b}). The reach of LHC for SUSY for various colliders assuming
100 fb$^{-1}$ at LHC and 10 fb$^{-1}$ at Tevatron (in the mSUGRA model).
}
\label{fig:lhc}
\end{figure}

It is convenient to classify events according to the isolated leptons:
\bi
\item $0\ell +\eslt +$jets
\item $1\ell +\eslt +$ jets
\item opposite-sign (OS) dileptons $+\eslt +$ jets
\item same-sign (SS) dileptons $+\eslt +$ jets
\item $3\ell +\eslt +$ jets,
\item $4\ell +\eslt +$ jets,
\item $\cdots$ .
\ei
One may simulate many thousands (millions) of signal events in SUSY model parameter
space, and compare against SM BG rates (from $t\bar{t}$, $W+$jets, $Z+$jets, $VV$, etc.
($V=W,\ Z$ or $\gamma$). For low mass SUSY particles, softer cuts work best, while for
high mass SUSY, hard cuts are needed. 
Typically, we generate a large grid of cuts, and optimize over all the various cut- channels.
Then one may see where a $5\sigma$/10 event signal is seeable for an assumed integrated
luminosity. An example of the LHC SUSY reach in the $m_0\ vs.\ m_{1/2}$ plane of the 
mSUGRA model is shown in Fig. \ref{fig:lhc}{\it a})., assuming 100 fb$^{-1}$ of integrated luminosity.
The furthest reach occurs in the $0\ell +$jets channel, and is $m_{\tg}\sim 3$ TeV
when $m_{\tq}\simeq m_{\tg}$ (left side of plot), or $m_{\tg}\simeq 1.8$ TeV for
$m_{\tq}\gg m_{\tg}$ (right side of plot)\cite{lhcreach}.

The LHC reach can be compared to the Tevatron reach (via clean trileptons), and
an ILC reach for $e^+e^-$ collisions at $\sqrt{s}=0.5$ and $1$ TeV in 
Fig. \ref{fig:lhc}{\it b}). Usually the LHC reach is dominant, except in the HB/FP region (right side of
plots). 
In this region, squarks and possible gluinos are very heavy, and will be produced at low cross sections. However, since $\mu$ is small, chargino production can occur at large
rates at ILC, and these should be easily visible. However, over most of parameter space,
LHC reach dominates. The main virtue of the ILC is that it can do high precision
sparticle spectroscopy, while the LHC is somewhat more of a discovery machine
(although it has good capability for sparticle mass reconstruction in regions
where simple production and decay modes dominate)\cite{frank}.

\section{Early discovery of SUSY at LHC via multi muons or dijets}

LHC is expected to turn on in November with $pp$ collisions at $\sqrt{s}\sim 7$ TeV, and to continue taking data
for about a year. A common strategy is to first ``re-discover'' the SM, and once that is well
understood, then begin the seach for new physics. Theorists are an impatient bunch, however, and
it is worth asking if early search and possible discovery of SUSY is possible during the
first year of LHC running. Jets should be easily visible, although some energy calibration is
necessary. 
Muons are easily identified, and in fact Atlas and CMS have already seen 
millions of cosmic muons, which have been used for alignment purposes. Electrons
will be identifiable, although at first there may be a larger-than-desired  $e$-jet
differentiation problem. The hardest thing is to well-measure $\eslt$ since $\eslt$ is the
negative of everything that is seen, and hence requires a {\it complete} knowledge of
the detector. Indeed, experience from turning on the D0 detector shows that multi-bump
$\eslt$ spectra may be expected until the detector becomes well-known, 
and the data is cleaned up.

\begin{figure}
  \includegraphics[height=.3\textheight]{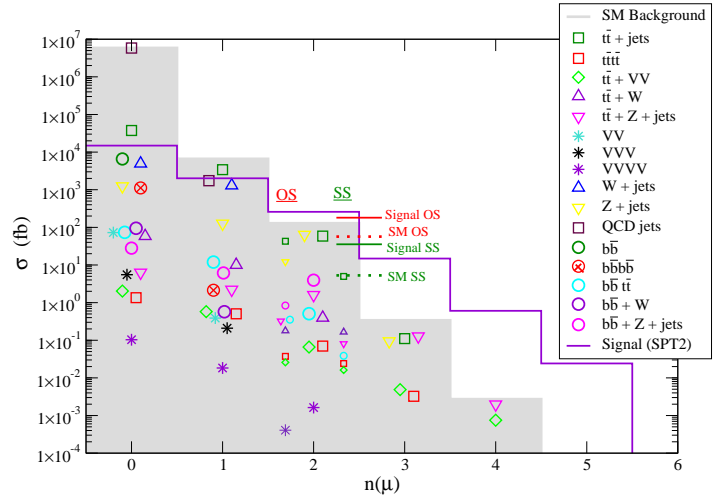}
  \caption{SUSY signal and SM BG for multi-muon plus $\ge 4$ jets events at 
$\sqrt{s}=10$ TeV LHC.
}
\label{fig:nmu}
\end{figure}

We examined if one can abandon the $\eslt$ cut, and use some other cut to elicit SUSY signal
from BG. The answer is {\it yes}, in the case of high isolated muon multiplicity\cite{early,bblt}.
We plot event rate for multi-muon plus $\ge 4$ jets events for signal 
(mSUGRA point (450,170,0,45,+1) purple histogram)
and summed SM background (gray histogram) in Fig. \ref{fig:nmu}.
We see that SM BG dominates signal in the $0\mu$ and $1\mu$ channels.
In the dimuon channel-- for both OS and SS dimuons-- signal now exceeds BG. In the trimuon
channel, signal exceeds BG by a factor of about 20! And this is without any $\eslt$ cut.

\begin{figure}
  \includegraphics[height=.3\textheight]{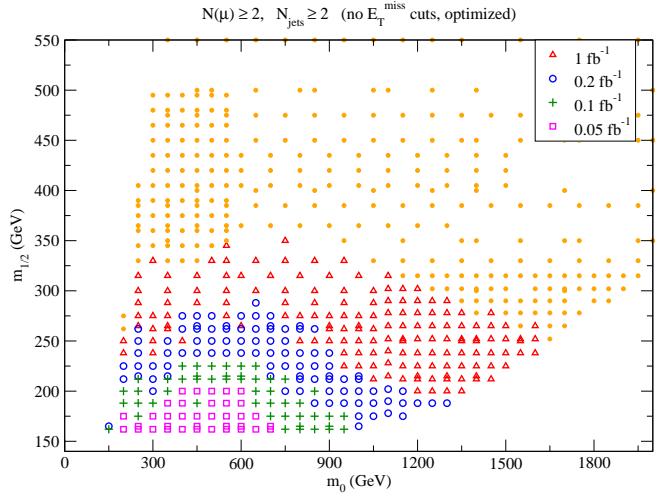}
  \caption{SUSY reach of LHC with $\sqrt{s}=10$ TeV for OS dimuons with optimized cuts
for various low levels of integrated luminosity.
}
\label{fig:mureach}
\end{figure}

The LHC reach for $\sqrt{s}=10$ TeV in multi-muon plus $\ge 4$ jets channel is shown in Fig.
\ref{fig:mureach}. Here, we see already some significant SUSY reach mainly in the OS dimuon channel for
integrated luminosity as low as $50$pb$^{-1}$\cite{bblt}! 
In the OS dimuon channel, in the favorable case,
we expect production of $\tz_2$ in gluino and squark cascade decays, followed by
$\tz_2\to\mu^+\mu^-\tz_1$. The dimuon mass distribution should build up a bump with kinematic cut-off
sitting right in between the photon and $Z$ poles. The signal should be readily apparent.

In addition to signals in the multi-mu plus jets channel, Randall and Tucker-Smith
propose looking for SUSY in the dijets channel\cite{rts}-- 
again without $\eslt$. Using judicious cuts,
especially $\Delta\phi (jj)$, indeed SUSY signal stands out from QCD BG at low integrated 
luminosity! The reach in this channel is best at low $m_0$, where squark pair production
occurs at large rates, leading to SUSY dijet events\cite{bblt}.

\section{Prospects for Yukawa unified SUSY at LHC}

Finally, let me address Yukawa-unified SUSY. At the beginning of this talk, I mentioned
that SUSY GUT theories based on the gauge group $SO(10)$ have a high degree of
motivation in that they unify matter, in addition to unification of
gauge couplings. The matter unification only works if one introduces a superfield
$\hat{N}^c_i$ ($i=1-3$ a generation index) containing a gauge singlet right hand neutrino state: 
exactly what is needed to give
neutrinos mass and to explain neutrino oscillation data. 
In the simplest models, one also expects the Yukawa couplings of the third generation to
unify at the GUT scale: $t-b-\tau$ unification.
A scan over parameter space of the mSUGRA model shows that such unification-- which depends
strongly on the $t$, $b$ and $\tau$ masses, and on their weak scale threshold corrections--
cannot occur\cite{abbbft}. At large $\tan\beta\sim 50$, they would like to unify, but
radiative EWSB breaks down because the down-Higgs soft term runs more negative than the 
up-Higgs soft term. A way around this is to move to the NUHM2 model where
$m_{H_d}^2>m_{H_u}^2$ at $M_{GUT}$, and neither is equal to $m_0$, the mSUGRA common scalar mass\cite{bf,bdr,bkss}.
This is called ``just-so'' Higgs splitting (HS). One can also use $SO(10)$ $D$-term splitting,
combined with generation splitting and inclusion of right-hand neutrino Yukawa 
coupling effects (the DR3 model\cite{dr3}). 

A scan over $SO(10)$ model parameter space reveals that indeed in the HS or DR3 model,
Yukawa unification can occur, but only for a very special range  SUSY parameters leading to
an inverted scalar mass hierarchy (IMH)\cite{bfpz}. 
The unification occurs if $m_{16}\sim 10$ TeV, and
$m_{1/2}$ is as small as allowed by LEP2 experiments. Also, $A_0^2=2m_{10}^2=4m_{16}^2$.
The specific parameter space leads to a specific sparticle mass spectrum, characterized by
\bi
\item first/second generation squarks and sleptons $\sim 10$ TeV,
\item third generation scalars and heavy Higgs and $\mu$ $\sim$ 1-2 TeV,
\item gluino mass $\sim 300-500$ GeV
\item light chargino $\sim 100-180$ GeV
\item a bino-like $\tz_1$ with $m_{\tz_1}\sim 50-80$ GeV.
\ei
The model is very compelling, except that it predicts a neutralino relic abundance
of around $10^2-10^4$ times that measured by WMAP. So it is excluded? No. 
Here, we invoke the PQWW solution to the strong CP problem, with an axino
mass $\sim$ MeV scale. Then the $\tz_1\to\ta\gamma$ decay reduces the relic abundance
by a factor of $10^3-10^5$! The cosmology was investigated in Ref. \cite{bhkss},
and works best if dark matter is composed of mainly cold axions, with a small contribution of
thermal and non-thermal axinos. We expect the gravitino mass $m_{\tG}\sim m_{16}\sim 10$ TeV,
which means the gravitino decays before the onset of BBN, so the gravitino problem is solved.
In addition, the model allows a re-heat temperature $T_R\sim 10^7$ GeV: enough to sustain
at least non-thermal leptogenesis (wherein the heavy right-hand neutrino states are produced via inflaton decay). 

The whole scenario is very compelling, but how do we test it? Well, the rather light gluinos
mean gluino pairs will be produced at high rates at the LHC, and may even yield signals
during the first year of running in the multi-mu plus jets channel, or multi-$b$-jet channel\cite{so10lhc}.
The dimuon spectrum should exhibit the characteristic mass edge around 
$m_{\tz_2}-m_{\tz_1}\sim 50-80$ GeV. Also, the prediction is that there will be no
signals in any direct or indirect WIMP search channels. However, the ADMX experiment stands
a good chance of finding an axion signal in their cryogenic microwave cavity experiment\cite{admx}.

\section{Conclusions}

I present my conclusions as a bullet list.
\bi
\item In many gravity mediated models, such as mSUGRA, the axion/axino mixture
is (IMO) a better candidate for CDM than neutralinos. 
If so, then entirely different regions of model parameter space are preferred.
\item Direct production of slepton pairs can be searched for at 
LHC up to $m_{\tell}\sim 350$ GeV for $\sim 10$ fb$^{-1}$ of integrated luminosity.
\item Clean trileptons from $\tw_1\tz_2$ production should be visible unless $\tz_2$
spoiler modes open up (or large interferences occur in 3-body decays).
\item Multi-lepton plus multi-jet$+\eslt$ events offer the best LHC reach for SUSY:
for 100 fb$^{-1}$, $m_{\tg}\sim 1.8-3$ TeV can be probed, depending on $m_{\tq}$.
\item It is possible to perform an early search for SUSY in multi-mu plus jets
{\it without} $\eslt$ channel (or dijet channel) even with very low integrated luminosity.
\item Yukawa unified SUSY has a very characteristic spectrum with light gluinos
and heavy squarks. It gives robust signatures at LHC in the first year of running.
It is cosmologically viable if the dark matter is composed of an axion axino
admixture (but with mainly axions). 
\ei






\begin{thebibliography}{9}
%
\bibitem{wss} For a review of SUSY, see
H.~Baer and X.~Tata, {\it Weak Scale Supersymmetry: From 
Superfields to Scattering Events}, 
(Cambridge University Press, 2006)
%
\bibitem{sugra} A.~Chamseddine, R.~Arnowitt and P.~Nath,
  Phys. Rev. Lett. {\bf 49}, 970 (1982); R.~Barbieri, S.~Ferrara and
  C.~Savoy, Phys. lett B {\bf 119}, 343 (1982); N.~Ohta,
  Prog. Theor. Phys. {\bf 70}, 542 (1983); L. Hall, J. Lykken and
  S. Weinberg, Phys. Rev. D {\bf 27}, 2359 (1983)
%
\bibitem{haim} H.~Goldberg, \prl{50}{1983}{1419};
J.~Ellis {\it et al.} \npb {238}{1984}{453}.
%
\bibitem{bb} H.~Baer and C.~Balazs, JCAP {\bf 0305} (2003) 006.
%
\bibitem{kkmy} M. Kawasaki, K. Kohri, T. Moroi and A. Yotsuyanagi, \prd{78}{2008}{065011}.
%
\bibitem{pqww}R. Peccei and H. Quinn, \prl{38}{1977}{1440} and
\prd{16}{1977}{1791}; S. Weinberg, \prl{40}{1978}{223};
F. Wilczek, \prl{40}{1978}{279}.
%
\bibitem{rtw}  K. Rajagopal, M. Turner and F. Wilczek, 
\npb{358}{1991}{447}.
%
\bibitem{ckkr} L. Covi, J. E. Kim and L. Roszkowski, \prl{82}{1999}{4180}; 
L. Covi, H. B. Kim, J. E. Kim and L. Roszkowski, \jhep{0105}{2001}{033}.
%
\bibitem{as} L. F. Abbott and P. Sikivie, \plb{120}{1983}{133};
J. Preskill, M. Wise and F. Wilczek, \plb{120}{1983}{127};
M. Dine and W. Fischler, \plb{120}{1983}{137};
M. Turner, \prd{33}{1986}{889}.
%
\bibitem{bs} A. Brandenburg and F.~Steffen,
JCAP{\bf 0408} (2004) 008. 
%
\bibitem{bbs} H. Baer, A. Box and H. Summy, arXiv:0906.2595 (2009).
%
\bibitem{ntlepto} G. Lazarides and Q. Shafi, \plb{258}{1991}{305};
K. Kumekawa, T. Moroi and T. Yanagida, \ptp{92}{1994}{437};
T. Asaka, K. Hamaguchi, M. Kawasaki and T. Yanagida, \plb{464}{1999}{12}.
%
\bibitem{bcpt1} H. Baer, C. H. Chen, F. Paige and X. Tata, \prd{49}{1994}{3283}.
%
\bibitem{dr} D. Denegri, W. Majerotto and L. Rurua, \prd{58}{1998}{095010}.
%
\bibitem{lhc3l} H. Baer, C. H. Chen, F. Paige and X. Tata, \prd{50}{1994}{4508}.
%
\bibitem{frank} I. Hinchliffe, F. Paige, M. Shapiro, J. Soderqvist and W. Yao,
\prd{55}{1997}{5520}.
%
\bibitem{cascade} H. Baer, J. Ellis, G. Gelmini, D. V. Nanopoulos
and X. Tata, \plb{161}{1985}{175};
G. Gamberini, \zpc{30}{1986}{605}; H. Baer, V. Barger,
D. Karatas and X. Tata, \prd{36}{1987}{96}.
%
\bibitem{ltanb} H. Baer, C. H. Chen, M. Drees, F. Paige and X. Tata,
\prl{79}{1997}{986}.
%
\bibitem{lhcreach} H.~Baer, C.~H.~Chen, F.~Paige and X.~Tata, \prd{52}{1995}{2746} and 
\prd{53}{1996}{6241}; 
H.~Baer, C.~H.~Chen, M.~Drees, F.~Paige and X.~Tata, \prd{59}{1999}{055014}
H.~Baer, C.~Bal\'azs, A.~Belyaev, T.~Krupovnickas and X.~Tata,
\jhep{0306}{2003}{054}; see also,  
S.~Abdullin and F.~Charles, \npb{547}{1999}{60};
S.~Abdullin {\it et al.} (CMS Collaboration), \jphg{28}{2002}{469};
B.~Allanach, J.~Hetherington, A.~Parker and B.~Webber, 
\jhep{08}{2000}{017}.
%
\bibitem{early} H. Baer, H. Prosper and H. Summy, \prd{77}{2008}{055017};
H. Baer, A. Lessa and H. Summy, \plb{674}{2009}{49}.
%
\bibitem{bblt} H. Baer, V. Barger, A. Lessa and X. Tata, arXiv:0907.1922 (2009).
%
\bibitem{rts} L. Randall and D. Tucker-Smith, \prl{101}{2008}{221803}.
%
\bibitem{abbbft} D. Auto, H. Baer, C. Balazs, A. Belyaev, J. Ferrandis 
and X. Tata, \jhep{0306}{2003}{023}.
%
\bibitem{bf} H. Baer and J. Ferrandis, \prl{87}{2001}{211803}.
%
\bibitem{bdr} T. Blazek, R. Dermisek and S. Raby, \prl{88}{2002}{111804} and \prd{65}{2002}{115004}.
%
\bibitem{bkss} H. Baer, S. Kraml, S. Sekmen and H. Summy, \jhep{0803}{2008}{056}.
%
\bibitem{dr3} H. Baer, S. Kraml and S. Sekmen, arXiv:0908.0134 (2009).
%
\bibitem{bfpz} J. Bagger, J. Feng, N. Polonsky and R. Zhang, \plb{473}{2000}{264}.
%
\bibitem{bhkss} H. Baer and H. Summy, \plb{666}{2008}{5}; 
H. Baer, M. Haider, S. Kraml,  S. Sekmen and H. Summy,
JCAP{\bf 0902} (2009) 002.
%
\bibitem{so10lhc} H. Baer, S. Kraml, S. Sekmen and H. Summy,
\jhep{0810}{2008}{079}.
%
\bibitem{admx} L. Duffy {\it et al.}, \prl{95}{2005}{091304} and \prd{74}{2006}{012006}.
%
\end{thebibliography}
\end{document}